\documentclass[amsmath,twocolumn,superscriptaddress,showpacs,aps,prb]{revtex4}
\usepackage{bm}
\usepackage{graphics}
\begin{document}
\title{Plasmon attenuation and optical conductivity of a two-dimensional electron gas}
\author{E.G.\ Mishchenko}
\affiliation{Lyman Laboratory, Department of Physics, Harvard
University, Cambridge, MA 02138} \affiliation{L.D. Landau
Institute for Theoretical Physics, Moscow 117334, Russia}
\author{M.Yu.\ Reizer}
\affiliation{5614 Naiche Rd., Columbus, OH 43213}
\author{L.I.\ Glazman}
\affiliation{Theoretical Physics Institute, University of
Minnesota, Minneapolis, MN 55455}

\begin{abstract}{
In a ballistic two-dimensional electron gas, the Landau damping
does not lead to plasmon attenuation in a broad interval of wave
vectors $q\lesssim k_F$. Similarly, it does not contribute to the
optical conductivity $\sigma (\omega, q)$ in a wide domain of
its arguments, $E_F>\omega>qv_F$, where $E_F$, $k_F$ and $v_F$ are,
respectively, the Fermi energy, wavevector and velocity of the
electrons. We identify processes that result in the  plasmon
attenuation in the absence of Landau damping. These processes are:
the excitation of two electron-hole pairs, phonon-assisted
excitation of one pair, and a direct plasmon-phonon conversion. We
evaluate the corresponding contributions to the plasmon linewidth
and to the optical conductivity.}
\end{abstract}

\maketitle

\section{Introduction}

Because of the long-range nature of the Coulomb interaction, the
energy of a plasmon propagating in an electron gas exceeds greatly
the energy of an electron-hole excitation at the same wave vector
$q$, provided that $q$ is smaller than the Fermi wave vector
$k_F$. This remains true in any dimensionality, including the case
of a two-dimensional electron gas (2DEG), for which the plasmon
spectrum is gapless,
\begin{equation}
\label{spectrum} \omega_q= v_F\sqrt{\frac{\kappa q}{2}}\,;
\end{equation}
here $v_F$ is the Fermi velocity, $\kappa=2 m e^2/\varepsilon^*$
is the inverse screening radius, $m$ is the effective electron
mass, and $\varepsilon^*$ is the dielectric constant of the host
material. (Hereinafter, we use the units with $\hbar =1$.) High
plasmon velocity, $d\omega_q/dq\gg v_F$, prevents plasmon from
decaying into an electron-hole pair and makes the Landau damping
exponentially small at low temperatures. Therefore, the leading
contribution to the plasmon attenuation of a purely electronic
nature comes from the plasmon decay into two electron-hole
pairs\cite{DBK}. Two such pairs propagating in opposite directions
can carry large energy while having a negligible total momentum.

In 2DEG formed in semiconductor heterostructures, there are
additional channels for plasmon attenuation. The energy and
momentum conservation conditions can also be satisfied for
processes in which a plasmon creates an electron-hole pair and a
phonon, rather than two electron-hole pairs. We will call an
attenuation due to such processes a {\it phonon-assisted Landau
damping} since it involves a single electron-hole pair. The
effectiveness of this process is increased by a large phase space
for the emission of phonons in the bulk of a semiconductor.

Yet another possibility of the plasmon attenuation in a
heterostructure arises from a conversion of the plasmon into an
acoustic phonon via a virtual electron-hole pair. Such a process
is feasible due to the absence of momentum conservation in the
direction perpendicular to the 2DEG plane ($z$-direction). This
makes it possible to satisfy energy and in-plane momentum
conservation conditions for a process in which a plasmon is
converted into a phonon with appropriate value $q_z$ of the
momentum along $z$-axis: $\omega_q=s\sqrt{q^2+q_z^2}$, where $s$
is the sound velocity. Note that processes involving optical
phonons can be safely neglected as they have large energy
thresholds (typically $35-50$ meV).

In the present paper, we analyze the mentioned above inelastic
processes leading to plasmon attenuation, and do not consider the
effect of impurities, assuming 2DEG sufficiently clean. In Section
II the plasmon broadening due to plasmon-phonon conversion is
discussed.  The conversion is mediated by a virtual electron-hole
pair with a rate independent of temperature. In subsequent
Sections we discuss scattering processes leading to the creation
of real electron-hole pairs in the final state: in Section III the
probability for plasmon scattering into a phonon and an
electron-hole pair is derived. In Section IV we evaluate the rate
of plasmon decay into two electron-hole pairs. In Section V we
compare the considered mechanisms with each other.

\section{Plasmon-phonon conversion}

The Hamiltonian of the 2DEG interacting with phonons has the form,
\begin{eqnarray}
\label{hamil} \hat{H}= - \sum_i\frac{\nabla^2_i}{2m}
+\sum_{i>j} \frac{e^2}{\varepsilon^*|{\bf x}_i-{\bf x}_j|} \nonumber\\
+\sum_{i \lambda} \int \frac{d^3k}{(2\pi)^3} \frac{\cal
M_{\lambda}({ \bf k})}{\sqrt{2\rho \omega_{\lambda \bf k}}}
\left(\hat c_{\lambda \bf
k}e^{-i\omega_{\lambda \bf k} t+i{\bf k}_\parallel {\bf x}_i} + \text{c.c.} 
\right),
\end{eqnarray}
where the second quantization representation for phonon variables and
coordinate representation for electron variables are used. The second
term in Eq.~(\ref{hamil}) stands for the electron-electron
interaction and the last term describes interaction of two-dimensional
electrons with three-dimensional phonons: ${\cal M}_\lambda({\bf k})$
is the coupling function for the $\lambda$-th phonon branch, $\hat
c_{\lambda \bf k}$ are the phonon annihilation operators, ${\bf
  k}_\parallel$ denotes the in-plane component of the phonon wave
vector ${\bf k}$, and $\rho$ is the density of the crystal. For
simplicity, we assume an isotropic phonon spectrum, $\omega_{\lambda
  \bf k}=s_\lambda k$.

The electron-phonon interaction in semiconductor heterostructures
is mainly due to the deformation potential (both Si- and
GaAs-based structures) and piezoelectric coupling (GaAs
structures). The deformation potential\cite{P,MPH,WRJ},
\begin{equation}
\label{defpot} {\cal M}^2_{l}({ \bf k})= \Xi^2 (
k_\parallel^2+k_z^2),
\end{equation}
couples electrons to the longitudinal ($l$) acoustic phonons only,
$\Xi$ being the deformation potential constant.

Piezoelectric interaction (present in GaAs heterostructures)
couples  electrons both to longitudinal and transverse ($t$)
phonons\cite{M,Z,BFS},
\begin{equation}
\label{piezol} {\cal M}^2_{\lambda}({\bf k}) = 2(e h_{14})^2
\frac{k_\parallel^2}{k^6} \left\{ \begin{array}{ll} \frac{9}{4}
k_\parallel^2k_z^2, & \lambda=l,\\
k_z^4+\frac{1}{8}k_\parallel^4, &\lambda=t.
\end{array} \right.
\end{equation}
Here we utilized the conventional notation for the coupling
constant, $h_{14}$. In contrast to the deformation potential
interaction Eq.\ (\ref{defpot}), piezoelectric coupling does not
vanish in the long-wavelength limit ${\bf k}\to 0$.

The plasmon dispersion is determined by the solution of the
equation $\varepsilon
(\omega,q)/\varepsilon^*=1-V_q\chi(\omega,{q})=0$. Here $V_q=2\pi
e^2/(q\varepsilon^*)$ stands for the Fourier transform of the
Coulomb interaction. The electron polarization operator in the
random phase approximation (RPA) is given by
\begin{equation}
\label{chi} \chi(\omega,{q})=\int\frac{2d^2p}{(2\pi)^2}
\frac{n_{{\bf p}-{\bf q}}-n_{{\bf p}}}{\omega +\xi_{{\bf p}-{\bf
q}}-\xi_{{\bf p}}+i\eta},
\end{equation}
where $n_{\bf p}$ is the Fermi-Dirac distribution function. In the
plasmon frequency range ($\omega \gg qv_F$) the polarization
operator can be approximated by $\chi(\omega,q)=m
q^2v_F^2/(2\pi\omega^2)$, leading to the plasmon spectrum,
Eq.~(\ref{spectrum}).

\begin{figure}[h]
\label{1} \resizebox{.46\textwidth}{!}{\includegraphics{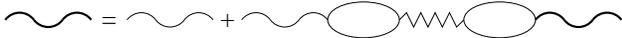}}
\caption{The Dyson equation for the propagator of the scalar
potential (shown by the bold wavy line) in the lowest order in the
electron phonon-interaction. The thin wavy line stands for the
propagator evaluated in the conventional RPA approximation,
$U(\omega,q)=V_q/[1-V_q\chi(\omega,q)]$, the zigzag line
represents the phonon Green function, the loop stands for the
electron polarization operator $\chi(\omega,q)$.}
\end{figure}

To find the lowest-order in electron-phonon interaction correction to
the dielectric function, one needs to solve the equation for the
longitudinal electric field propagator, see Fig.~1. The resulting
dielectric function is
\begin{eqnarray}
\label{impr_RPA} \varepsilon (\omega,{q})/\varepsilon^* &=& 1 -
V_q\chi(\omega,{ q})-V_q\chi^2(\omega,{q})\nonumber\\ & \times &
\sum_{\lambda}\int\frac{dq_z}{2\pi} \frac{{\cal M}_\lambda^2 ({
q},q_z)}{\rho[(\omega+i\eta)^2-s^2_{\lambda}(q^2+q_z^2)]}.~~
\end{eqnarray}
 The plasmon spectrum
$\omega=\omega_q-i\gamma_q$  is determined by a zero of the
dielectric function (\ref{impr_RPA}). The imaginary part
$\gamma_q$ originates from the poles of the phonon propagator at
those values of momentum $q_z$ which satisfy the energy
conservation condition, $\omega_q=s_\lambda \sqrt{q^2+q_z^2}$.
After simple calculation, with the use of the condition $\vert q_z
\vert \approx \omega_q/s_\lambda \gg q$, we obtain
\begin{equation}
\label{homowidth} \gamma_q=\frac{m q^2 v_F^2}{8\pi \rho
\omega_q^2} \sum_\lambda s_\lambda^{-1} {\cal M}_\lambda^2
(q,\omega_q/s_\lambda).
\end{equation}
The deformation potential interaction yields the contribution to
the plasmon linewidth,
\begin{equation}
\label{homowidth_d} \gamma_q=\frac{\pi \eta_d^2}{4}
\frac{q^2v_F^2}{\omega_{s}}.
\end{equation}
Here $\omega_{s} =k_F s_l$ is the characteristic phonon frequency,
and
\begin{equation}
\label{coupling_d} \eta_d^2=\frac{mk_F\Xi^2}{2\pi^2\rho s_l^2 }
\end{equation}
is the dimensionless coupling constant for the deformation
potential interaction.

The piezoelectric coupling, contrary to the deformation potential,
remains finite in the long-wavelength limit. Therefore, when present,
the former often leads to much stronger effects than the latter.
However, the piezoelectric contribution into the width
(\ref{homowidth}) is strongly reduced due to the fact that the
plasmons emit phonons almost perpendicular to the 2DEG plane, $\theta
\approx q s/\omega_q \ll 1$. According to Eq.~(\ref{piezol}), the
piezoelectric coupling is stronger for the transverse phonons.
However, the corresponding contribution to the width
Eq.~(\ref{homowidth}) is still small compared to the deformation
potential contribution Eq.~(\ref{homowidth_d}), by a factor of
$(s/v_F)^4 \ll 1$.

Rate of the direct (via a virtual electron-hole pair)
plasmon-phonon conversion given by Eq. (\ref{homowidth}),
describes the temperature-independent plasmon broadening. To
analyze the temperature-dependent attenuation one has to account
for the inelastic processes resulting in creation of electron-hole
pairs in the final state. To do this we relate the plasmon
attenuation to the optical conductivity.

\section{Phonon-assisted Landau damping}

The propagation of a plasmon in 2DEG is accompanied by an
oscillating electric field, whose scalar potential is a plane wave
in the in-plane direction ${\bf x}$ and satisfies the equation
$\nabla^2 \phi =0$ outside 2DEG,
\begin{equation}
\label{phi} \phi ({\bf x},z,t) = (\phi_0 e^{-i\omega t +i{\bf q
x}}+ \phi_0^* e^{i\omega t -i{\bf q x}})e^{-q|z|}.
\end{equation}
The energy (per unit area) of the electric field (\ref{phi}) is
given by
\begin{equation}
\label{energy} W_e= \frac{\varepsilon^*}{8\pi} \int\limits dz~
\overline{(\nabla \phi)^2}=\frac{\varepsilon^* q}
{2\pi}|\phi_0|^2={e^2}{V^{-1}_q} |\phi_0|^2,
\end{equation}
where the bar denotes the in-plane average.  The kinetic energy of
electrons is equal to the energy of electric field (the virial
theorem). The total energy of a plasmon is therefore, $W=2W_e$.
The plasmon attenuation $\gamma_q$ is related to the energy
dissipation rate,
\begin{equation}
\label{rate_w}
 {dW}/{dt}=-2\gamma_q W=-4\gamma_q W_e.
\end{equation}
On the other hand dissipation is related to the real part of the
longitudinal optical conductivity $\sigma'(\omega,{ q})$,
\begin{equation}
\label{sigma} -\frac{dW}{dt}=2 q^2 |\phi_0|^2 ~ \sigma'(\omega,{
q}).
\end{equation}
The right-hand side in this expression is the Joule heating. Note,
that Eq.~(\ref{sigma}) can be applied to attenuation of plasma
eigenmodes as well as the dissipation of the external electric
field.  In the latter case $q\phi_0$ should be understood as the
amplitude of the electric field at the 2DEG plane. Using
Eqs.~(\ref{rate_w}) and (\ref{sigma}) we can relate the plasmon
attenuation to the optical conductivity at $\omega=\omega_q$,
\begin{equation}
\label{gammaq} \gamma_q= \frac{q^2 V_q}{2e^2} ~ \sigma'(\omega_q,{
q}).
\end{equation}
Equation (\ref{gammaq}) may be viewed as the Ward identity
relating $\chi(\omega,q)$ and $\sigma(\omega,q)$, {\it i.e.}, the
polarization operators with scalar and vector vertices.

To evaluate $\gamma_q$, we find first $dW/dt$ with the help of the
perturbation theory in the interaction of electrons with phonons
and with electric field $\phi_0$, see Eqs.~(\ref{hamil}) and
(\ref{phi}). We then obtain $\gamma_q$ from $dW/dt$ with the help
of Eqs.~(\ref{sigma})--(\ref{gammaq}). We are interested in the
dissipation of the high-frequency electric field, $\omega \gg
qv_F$. At such frequencies, the field absorption due to an
excitation of a single electron-hole pair (Landau damping) is
forbidden by the energy and momentum conservation conditions. The
lowest-order absorption process, therefore, includes also a
creation or annihilation of a phonon, see Fig.~2.
\begin{figure}[h]
\resizebox{.40\textwidth}{!}{\includegraphics{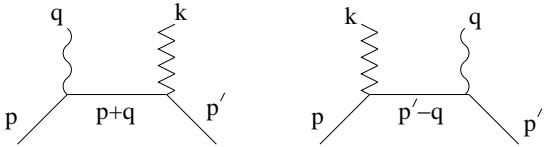}}
\caption{Graphic representation of the transition amplitude $L$.
Wavy lines stand for the amplitude of the electric field $\phi_0$,
zigzag lines denote phonons.  }
\end{figure}
The probability of the field absorption process is given by the
Fermi Golden rule\cite{LL},
\begin{eqnarray}
\label{trphonon} dw_{\pm }= 2\pi |L_\lambda|^2 \delta (\xi_{\bf
p}+ \omega-\xi_{\bf p'}\mp \omega_{\lambda \bf k})
\nonumber\\
\times \delta({\bf p}+{\bf q}-{\bf p'} \mp {\bf k}_\parallel)
\frac{d^2 p' d^3 k}{(2\pi)^3}.
\end{eqnarray}
Here the matrix element
\begin{eqnarray}
\label{matphonon} L_\lambda &=& \frac{e\phi_0\varepsilon^*~ \cal
M_{\lambda}({ \bf k})}{\varepsilon(\omega,k_\parallel)\sqrt{2\rho
\omega_{\lambda k}}}\nonumber\\ & \times & \left(
\frac{1}{\xi_{\bf p}-\xi_{\bf p+ q} + \omega }+\frac{1}{\xi_{\bf
p'}-\xi_{\bf p'- q} - \omega} \right)
\end{eqnarray}
accounts for the intermediate state in the process, and the signs
$\pm$ correspond to the creation (annihilation) of a phonon, which
accompanies the excitation of the electron-hole pair. The
dielectric function in the denominator of the matrix element
Eq.~(\ref{matphonon}) accounts for the screening of the
electron-phonon interaction by two-dimensional electrons. The
presence of two terms in Eq.~(\ref{matphonon}) corresponds to the
two different possibilities for the virtual state depending on
whether the electron interacts first with the plasmon or with the
phonon, see Fig.~2. Each internal line contributes to the matrix
element a propagator of the virtual state, $(E-E_v)^{-1}$, with
$E$ being the total energy of particles, and $E_v$ the energy of
the virtual state.

The resulting probability of the field absorption is given by the
appropriate sum over the initial and final states of the
electron-phonon system,
\begin{equation}
\label{ia} I_{a}=\sum_\lambda \int \frac{2d^2 p}{(2\pi)^2}
n_{p}(1-n_{p'})[(1+N_{\lambda k})dw_{+}+N_{\lambda k}dw_{-}],
\end{equation}
where $N_{\lambda k}$ is the Bose-Einstein phonon distribution,
and  factor $2$ takes into account the electron spin degeneracy.
The  probability for the field emission, $I_e$ can be found from
the detailed balance principle\cite{LLVIII},
$I_{e}=I_{a}e^{-\omega/T}$.

The energy dissipation rate is then determined by the two
probabilities,
\begin{equation}
\label{dissipphon1} -\frac{d W}{dt}=\omega (I_{a} -I_{e}) =\omega
(1-e^{-\omega/T})I_{a}.
\end{equation}
In the high-frequency domain $\omega \gg qv_F$, the relevant
momentum of an electron-hole pair ${\bf p}-{\bf p'}$ and that of a
phonon ${\bf k}$ are large compared to the plasmon momentum ${\bf
q}$, which allows us to simplify the matrix element $L_\lambda$,
\begin{equation}
\label{matphonon1} L_\lambda =  \frac{e\phi_0\varepsilon^*\cal
M_{\lambda}({ \bf k})}{\varepsilon(\omega,k_\parallel)\sqrt{2\rho
\omega_{\lambda k}}} \frac{({\bf p-p'})\cdot {\bf q}}{m\omega^2}.
\end{equation}
In the last expression we also have used the momentum conservation
condition Eq.\ (\ref{trphonon}). Combining Eqs.~(\ref{sigma}),
(\ref{trphonon}) and (\ref{ia})--(\ref{matphonon1}), we obtain the
phonon-assisted Landau damping in the form:
\begin{eqnarray}
\label{ph2deg}  \sigma' (\omega,q) = \frac{e^2 \varepsilon^{*2}
\sinh{[\frac{\omega}{2T}]}} {8\rho m^2 \omega^3}
 \sum_{\lambda}\int \frac{d^3k}{(2\pi)^3}\frac{k_\parallel^2 {\cal M}^2_{\lambda}({\bf
 k})}{ \omega_{\lambda k}\vert\varepsilon(\omega,k_\parallel)
 \vert^2} \nonumber\\
\times \frac{1}{\sinh{[\frac{\omega_{\lambda k}}{2T}]}} \Bigl[
  \frac{
  \chi''(\omega_{\lambda k}-\omega,{ k}_\parallel)}
{ \sinh{[\frac{\omega-\omega_{\lambda k}}{2T}]}}-\frac{
\chi''(\omega_{\lambda k}+\omega,{ k}_\parallel)} {
\sinh{[\frac{\omega+\omega_{\lambda k}}{2T}]}} \Bigr].
\end{eqnarray}
Here $\chi''$ is the structure factor of 2DEG, {\it i.e.}, the
imaginary part of the polarization operator Eq.~(\ref{chi}). We
are interested in the optical conductivity at $\omega, T\ll E_F$,
where $E_F=k_F^2/2m$ is the Fermi energy. It allows us to use the
low-frequency limit of $\chi''$,
$$
\chi''(\Omega
,{k}_\parallel)= -\frac{m}{\pi|k_\parallel| v_F}
\frac{\Omega 
}{ \sqrt{
1-(\frac{k_\parallel}{2k_F})^2}}.
$$
This limiting form is valid across the entire particle-hole domain
of excitations, except narrow regions near its ends,
$|\Omega|/v_F\leq k_\parallel\leq 2k_F-|\Omega|/v_F$.

The electron-phonon interaction is effectively screened at a small
momentum transfer between the two subsystems, which results in a
strong frequency dependence of the conductivity. Indeed, since
only the particle-hole domain is relevant in the integral
(\ref{ph2deg}), it suffices to approximate the dielectric function
by its static limit,
$\varepsilon(0,k_\parallel)/\varepsilon^*=1+\kappa^2/k^2_\parallel$.
The characteristic momenta in the integral of Eq.~(\ref{ph2deg})
are $k_\parallel\sim \text{max}(\omega,T)/s$. Therefore, at
$\omega,T \ll \omega_\kappa$ the screening of the electron-phonon
interaction is strong (here we introduced the characteristic
frequency $\omega_\kappa=\kappa s$, and $\kappa$ is the inverse
screening radius). We consider below the plasmon attenuation in
the limits of weak and strong screening.

(i) In the limit of low frequency and temperature, $\omega,T \ll
\omega_\kappa$, corresponding to the strong screening, a
straightforward evaluation of the integral in Eq.~(\ref{ph2deg})
yields for the electron-phonon interaction via deformation
potential,
\begin{equation}
\label{def2} \gamma_q  = \frac{3\pi\eta_d^2 } {7\cdot
2^9~\omega_s^4\omega_\kappa^2} \left\{
\begin{array}{ll}\omega_q^7, & \omega_q > 2\pi T,\\
A (2\pi T)^7, & \omega_q < 2\pi T, \end{array} \right.
\end{equation}
where $A=1.47$. Similarly, for the piezoelectric coupling,
\begin{equation}
\label{gammapiezo2} \gamma_{q}=\frac{\pi\eta_p^2}{15\cdot 2^5~
\omega_s^2 \omega^2_\kappa} \left\{
\begin{array}{ll} \omega_q^5,& \omega_q >2\pi T,\\
B (2\pi T)^5,& \omega_q<2\pi T, \end{array} \right.
\end{equation}
with the coefficient $B=0.76$. Here characteristic frequencies
$\omega_\kappa =\kappa s_l$ and $\omega_s=k_Fs_l$ depend on the
material parameters.  The deformation potential coupling strength
$\eta_d$ is given in Eq.\ (\ref{coupling_d}), and the
dimensionless constant for the piezoelectric coupling is
\begin{equation}
\label{piezoconstant} {\eta}_p^2=\frac{m (eh_{14})^2}{2\pi^2 \rho
s_l^2 k_F} \left(\frac{63}{256} +\frac{s_l^6}{s_t^6}
\frac{87}{256} \right).
\end{equation}
Clearly, in the limit of low temperature and plasmon frequency,
the piezoelectric part, Eq.~(\ref{gammapiezo2}), prevails over the
deformation potential contribution Eq.\ (\ref{def2}).

(ii) We turn now to the case of a weak screening. In typical
heterostructures the inverse screening radius $\kappa$ is of the
order of $k_F$, and therefore $\omega_\kappa \sim \omega_s$. We
concentrate on relatively large frequencies or high temperatures,
 {\it i.e.}, assume that $\omega$ or $T$ exceed $\omega_\kappa$ and
$\omega_s$, but are still small compared to the Fermi energy,
$E_F=k_F^2/2m$.

The leading deformation potential contribution in
Eq.~(\ref{ph2deg}) comes from phonon states with $k_\parallel \sim
k_F$ and $k_z\sim \text{max} (\omega,T)/s \gg k_F$, {\it i.e.}, from the
phonons propagating almost normally to the 2DEG plane. In the
limit of high temperature or frequency we find
\begin{equation}
\label{def1} \gamma_q  = C\frac{\eta_d^2} {\omega_s}
\bigl[\omega_q^2+(2\pi T)^2\bigr].
\end{equation}
Here the numerical coefficient $C$ depends on the screening
parameter $\kappa/2k_F$,
\begin{equation}
C=\frac{1}{3}\int\limits_0^{\pi/2}
\frac{\sin^4{\theta}}{(\sin{\theta}+\frac{\kappa}{2k_F})^2},
\end{equation}
and varies from $C=\pi/12\approx 0.26$ to $C=0.06$ when the
screening parameter increases from $\kappa/2k_F=0$ to
$\kappa/2k_F=1$.

Unlike the interaction via deformation potential, the
piezoelectric mechanism of the electron-phonon interaction is
ineffective for phonons with $|k_\parallel|\ll k$. As the result,
the piezoelectric mechanism yields a contribution to $\gamma_q$,
which is smaller than the one given by Eq.~(\ref{def1}). In
particular, at low temperatures and high frequencies,
$T\lesssim\omega_s\lesssim\omega$, this contribution to the
plasmon line width saturates at a value
\begin{equation}
\label{gammapiezo1} \gamma_q   \sim \eta_p^2~\omega_s,
\end{equation}
while in the case of high temperature, $T\gtrsim\omega_s$, we find
\begin{equation}
\label{gammapiezo3} \gamma_q   \sim \eta_p^2~ T,
\end{equation}
independent of the parameter $\omega/\omega_s$. Strictly speaking,
the piezoelectric constants in
Eqs.~(\ref{gammapiezo1})--(\ref{gammapiezo3}) differ from the
definition Eq.~(\ref{piezoconstant}) by the numerical coefficients in
the brackets, since the angle integrals are different in each
regime. However, for rough estimates, one can use
Eq.~(\ref{piezoconstant}).

\section{Two-pair absorption}

Another mechanism effective for the absorption of plasmons with
energies ($\omega_q \ll E_F$) is associated with creation of two
electron-hole excitations\cite{DBK}. The probability of absorption
or excitation of two electron-hole pairs by an electric field
Eq.~(\ref{phi}) can be found with the help of the Golden rule by
treating the field strength and the electron-electron interaction
$V_q$ perturbatively,
\begin{eqnarray}
\label{transition} dw
= 2\pi \vert {\cal L}\vert^2 \delta (\xi_{\bf p}+\xi_{\bf
k}-\xi_{\bf p'}-\xi_{\bf k'} + \omega)\nonumber\\ \times
\delta({\bf p}+{\bf k}-{\bf p'}-{\bf k'} + {\bf q}) \frac{d^2p'
d^2k'}{(2\pi)^2}.
\end{eqnarray}
Here ${\bf p}, {\bf k}$ and ${\bf p',k'}$ are, respectively, the
initial and final momenta of two electrons. The matrix element
${\cal L}$ depends of the initial spin state of the two electrons.
For a singlet state the transition matrix element is
\begin{widetext}
\begin{equation}
\label{updown} {\cal L}_{0,0} =e\phi_0\left(\frac{V_{\bf
k-k'}+V_{\bf p'-k}}{\xi_{\bf p}-\xi_{\bf p+ q} + \omega
}+\frac{V_{\bf k-k'}+V_{\bf p-k'}}{\xi_{\bf p'}-\xi_{\bf p'- q}-
\omega}+\frac{V_{\bf p'-p}+V_{\bf p-k'}}{\xi_{\bf k}-\xi_{\bf k +
q} + \omega} +\frac{V_{\bf p'-p}+V_{\bf p'-k}}{\xi_{\bf
k'}-\xi_{\bf k'- q}- \omega }\right),
\end{equation}
and for a triplet state it is
\begin{equation}
\label{upup} {\cal L}_{1,0} = e\phi_0\left(\frac{V_{\bf
k-k'}-V_{\bf p'-k}}{\xi_{\bf p}-\xi_{\bf p+ q} + \omega
}+\frac{V_{\bf k-k'}-V_{\bf p-k'}}{\xi_{\bf p'}-\xi_{\bf p'- q}-
\omega}+\frac{V_{\bf p'-p}-V_{\bf p-k'}}{\xi_{\bf k}-\xi_{\bf k +
q} + \omega} +\frac{V_{\bf p'-p}-V_{\bf p'-k}}{\xi_{\bf
k'}-\xi_{\bf k'- q}- \omega }\right).
\end{equation}
\end{widetext}
The first subscript of the matrix element denotes the value of the
total spin of two electrons, while the second subscript stands for
its projection onto the $z$-direction. The matrix element in the
triplet channel is the same for all three spin states, ${\cal
L}_{1,0}={\cal L}_{1,\pm 1}$, since the electron-electron
interaction is spin-independent. The structure of the matrix
elements Eqs.~(\ref{updown}) and (\ref{upup}) is clear from the
graphic representation shown in Fig.~3. Again, as in Fig.~2, the
internal lines correspond to the virtual states whose propagators
are given by $(E-E_v)^{-1}$, with $E$ being the total energy of
the particles and $E_v$ is the energy of a virtual state. The
total probability of the field absorption is given by the sum over
the initial and final states according to
\begin{equation}
\label{dissip} I_{a}=  \frac{1}{4} \int \frac{d^2p d^2k}{(2\pi)^4}
[dw_{0,0}+3dw_{1,0} ] n_{p}n_{ k}(1-n_{p'})(1-n_{k'}).
\end{equation}
Here the coefficient ${1}/{4}$ prevents from double-counting of
the initial and final states.
\begin{figure}[h]
\resizebox{.40\textwidth}{!}{\includegraphics{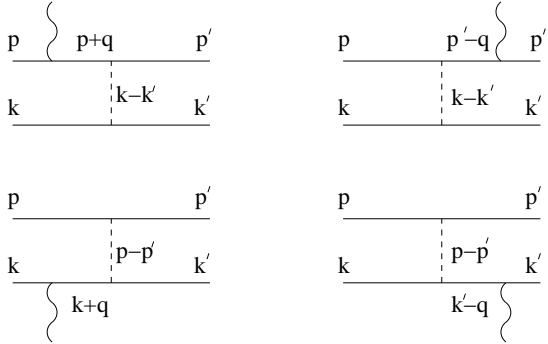}}
\caption{ Graphic representation for the two-pair absorption
amplitude ${\cal L}$, see Eqs.\ (\ref{updown})-(\ref{upup}). The
insertion of the wavy line (external wave or plasmon) into the
two-electron scattering process is possible in the shown here four
different ways, corresponding to four fractions in Eqs.\
(\ref{updown})-(\ref{upup}). In addition, an interchange ${\bf p'}
\leftrightarrow {\bf k'}$ in the final state is possible, bringing
two terms in each of the numerators in
Eqs.~(\ref{updown})-(\ref{upup}). The relative signs of these
terms depend on the spin state of the two electrons. }
\end{figure}
The above formulas are valid for any values of $\omega$ and $q$. We
now make use of the fact that $\omega \gg qv_F$ in the region of
interest, and obtain
\begin{eqnarray}
\label{st}
{\cal L}_{0,0} =\frac{e\phi_0}{m\omega^2}({\cal A} +{\cal A}_{\rm ex}),\nonumber\\
{\cal L}_{1,0} =\frac{e\phi_0}{m\omega^2}({\cal A} -{\cal A}_{\rm
ex}).
\end{eqnarray}
Here we introduced the following notations:
\begin{eqnarray}
\label{triplet} && {\cal A} = {\bf q }\cdot [({\bf Q+q})V_{\bf
Q+q}-{\bf Q}V_{ \bf Q}]\nonumber\\
&& {\cal A}_{\rm ex} = {\bf q}\cdot[({\bf p'-k})V_{\bf p'-k}
+({\bf k'-p})V_{\bf p-k'}],
\end{eqnarray}
and accounted for the momentum conservation evident from
Eq.~(\ref{transition}). The momentum transferred in a collision is
denoted by ${\bf Q}={\bf p}-{\bf p'}$. The term ${\cal A}_{\rm
ex}$ is obtained from ${\cal A}$ by interchanging the momenta
${\bf p'} \leftrightarrow {\bf k'}$, and originates from the
indistinguishability of colliding particles.

The transition rates given by Eq.~(\ref{transition}) with the
matrix elements Eq.~(\ref{st}) determine the total probability of
absorption Eq.~(\ref{dissip}). As usual, the total probability of
emission $I_e$ is related to the probability of absorption by the
detailed balance principle, $I_e=I_ae^{-\omega/T}$. The energy
dissipation rate $dW/dt$ can be expressed in terms of the
probability $I_a$ according to Eq.\ (\ref{dissipphon1}). The
optical conductivity is then obtained from its relation to
 $dW/dt$, represented by Eq.~(\ref{sigma}),
\begin{eqnarray}
\label{dissipnew} \sigma'(\omega,q)=
\frac{e^2(1-e^{-{\omega}/{T}})}{ m^2\omega^3
q^2} \int \frac{d^2p d^2k d^2p' d^2k'}{(2\pi)^{5}} \nonumber\\
\times {\cal A}\left ({\cal A}-{{\cal A}_{\rm ex}}/{2}
\right)\delta
(\xi_{\bf p}+\xi_{\bf k}-\xi_{\bf p'}-\xi_{\bf k'} + \omega)\nonumber\\
 \times
 n_{p}n_{ k}(1-n_{p'})(1-n_{k'}) \delta({\bf p}+{\bf k}-{\bf p'}-{\bf k'} + {\bf q}).
\end{eqnarray}

The optical conductivity Eq.~(\ref{dissipnew}) vanishes as
$\sigma'(\omega,q) \propto q^2$ in the long-wavelength limit $q\to 0$.
This is clear from the small--$q$ properties of the amplitudes ${\cal
  A}$ and ${\cal A}_{\rm ex}$. Indeed, the relation ${\cal A}\propto
q^2$ at $q\to 0$ follows directly from the first line of
Eq.~(\ref{triplet}), and the relation ${\cal A}_{\rm ex} \propto
q^2$ can be obtained from the second line of Eq.~(\ref{triplet})
and the momentum conservation condition. The absence of absorption
in the long-wavelength limit, $\sigma'(\omega,q=0)=0$, is a
manifestation of a more general principle\cite{PN}: a
translationally invariant electron liquid does not absorb energy
from an applied uniform ac field. From a technical standpoint, the
zero value of $\sigma'(\omega,q=0)$ comes as the result of a
cancellation of the four matrix elements, Fig.~3, of which any
element is not small by the external wave vector $q$. Recently,
Gornyi and Mirlin\cite{GM} demonstrated a similar cancellation in
the diagrammatic calculation of the homogeneous conductivity,
while Pustilnik et.\ al.\ encountered $q^2$-behavior in the
calculation of the far tails of the density-density correlation
function in the case of spinless one-dimensional
fermions\cite{PMA}. This cancellation was overlooked by Reizer and
Vinokur in their calculation of the attenuation of two-dimensional
plasmons\cite{RV}. It led to a result large by a factor
$k_F^2/q^2$, in contradiction with the above mentioned general
principle.

The contribution of the exchange term ${\cal A}_{\rm ex}$ to the
conductivity Eq.~(\ref{dissipnew}) can be estimated from
Eq.~(\ref{triplet}). This contribution involves large transferred
momenta $|{\bf p'-k}| \sim k_F$, whereas the small-momentum domain
of ${\bf Q} ={\bf p-p'}$ is important for the contribution of
${\cal A}$. Typical values of $Q$ in Eq.~(\ref{dissipnew}) can be
estimated from the conservation of energy, $Q \sim
\text{max}(\omega,T)/v_F$. From Eq.~(\ref{triplet}) we then obtain
${\cal A}_{\rm ex}/{\cal A} \sim V_Q/V_{k_F}$. The amplitude
${\cal A}_{\rm ex}$ can thus be neglected in
Eq.~(\ref{dissipnew}), if the interaction potential is
long-ranged, $V_{Q} \ll V_{k_F}$. This condition is met for
Coulomb interaction at $\omega, T \ll E_F$.  In this case the
scattering amplitudes are the same for singlet and triplet states.

Neglecting the exchange contribution one can conveniently
represent the optical conductivity in terms of the structure
factors for free electrons,
\begin{eqnarray}
\label{heat} \sigma'(\omega,{q})= \frac{e^2
\sinh{[\frac{\omega}{2T}]}}{2\pi  m^2 \omega^3q^2}
\int\frac{d^2Q}{(2\pi)^2}
 \bigl[{\bf q} \cdot({\bf Q}+ {\bf
q})V_{{\bf Q}+{\bf q}} \nonumber \\  -{\bf q}\cdot{\bf Q}V_{\bf
Q}\bigr]^2 \int\limits_{-\infty}^\infty d\Omega
~\frac{\chi''(\Omega,{\bf
 Q})
\chi''(\Omega+\omega,{\bf Q}+{\bf
q})}{\sinh{[\frac{\Omega}{2T}]}\sinh{[\frac{\Omega+\omega}{2T}]}}.
\end{eqnarray}
The formula (\ref{heat}) describes energy dissipation due to the
decay into two electron-hole pairs with energies and momenta
$\Omega+\omega,{\bf Q}+{\bf q}$ and $-\Omega,-{\bf Q}$,
respectively. The densities of the electron-hole pairs are
determined by the structure factors $\chi''$.

In the case of Coulomb interaction, $[{\bf q}\cdot ({\bf Q}+{\bf
q})~V_{{\bf Q}+{\bf q}}-{\bf q}\cdot{\bf Q}~V_{\bf Q}]^2 \propto
q^4/Q^2$. The momentum integral in Eq.~(\ref{heat}) is, therefore,
singular at small values of $Q$. If the smallest value of
$Q=\omega/2v_F$, allowed by the energy conservation condition,
becomes smaller than the inverse screening radius $\kappa$, then
the screening of Coulomb interaction must be accounted for. This
is achieved by the replacement of the interaction potential $V_Q$
in Eq.~(\ref{heat}) by its screened value. As in the case of the
phonon-assisted conductivity, it is sufficient here to account for
the dielectric function by using its static limit, $V_{Q} \to
\varepsilon^*V_{Q}/\varepsilon(0,Q)$. An emphasis on the small-$Q$
domain comes also from the structure of the density of states of
the electron-hole pairs. Indeed, integration over frequencies
$\Omega$ in Eq.~(\ref{heat}) results in an additional factor
$\propto Q^{-2}$. Therefore, if the calculation is to be performed
with the logarithmic in $\kappa _F/\omega$ accuracy, we can
neglect the difference between $V_{Q+q}$ and $V_Q$, and simplify
Eq.~(\ref{heat}) to
\begin{eqnarray} \label{attenuation1} \sigma'(\omega,q) =\frac{e^2
q^2 \sinh{[\frac{\omega}{2T}]}}{\pi m^2\omega^3}
\int\limits_0^\infty\frac{QdQ}{2\pi}\int\limits_{-\infty}^\infty
d\Omega
  \nonumber\\
\times \frac{V_Q^2~\varepsilon^{*2}}{\varepsilon^2(0,Q)} \frac{
\chi''(\Omega_+,{Q}) \chi''(\Omega_-,
Q)}{\cosh{[\frac{\Omega}{T}]}-\cosh{[\frac{\omega}{2T}]}},
\end{eqnarray}
where $\Omega_\pm =\Omega \pm \omega/2$. In the long-wavelength
limit, $Q\ll k_F$, the structure factor has the form
\begin{equation}
\label{imchi0} \chi''(\Omega,Q) = - \frac{m \Omega}{\pi
\sqrt{Q^2v^2_F-\Omega^2}} ~\theta (Qv_F-|\Omega|).
\end{equation}
To evaluate the conductivity Eq.~(\ref{attenuation1}), it is
technically more convenient to change the order of integrations
and evaluate first the integral over the momentum. With the
logarithmic accuracy, the integral is cut at the upper limit
$Q\sim \kappa$, leading to the following integral over frequency,
\begin{eqnarray}
\label{gamm} \sigma'(\omega ,q) &=& \frac{e^2
q^2\sinh{[\frac{\omega}{2T}]}~}{\pi^2 k_F^2 \omega^3}  \nonumber
\\  & \times & \int\limits_{0}^\infty d\Omega~ \frac{
(\frac{\omega}{2})^2-\Omega^2}
{\cosh{[\frac{\omega}{2T}]}-\cosh{[\frac{\Omega}{T}]}}
\ln{\bigl[\frac{ \kappa v_F }{\sqrt{\Omega\omega}}\bigr]}.~~~
\end{eqnarray}
Utilizing the fact that the logarithm is a slow function of its
argument, one can evaluate the integral in Eq.~(\ref{gamm}) with
the logarithm assigned its value at the characteristic frequency
$\Delta=\max(\omega,2\pi T)$. The two-pair contribution to the
optical conductivity takes the form
\begin{equation}
\label{twopaircon} \sigma'(\omega,q) = \frac{e^2q^2}{12\pi^2
k_F^2\omega^2} \left[ {\omega^2} + {(2\pi T)^2
}\right]\ln{\bigl[\frac{\kappa v_F}{\sqrt{\omega \Delta}}\bigr]}.
\end{equation}
Finally, the plasmon attenuation is found from its relation
Eq.~(\ref{gammaq}) to the optical conductivity,
\begin{equation}
\label{omega} {\gamma_q} = \frac{q^2}{24\pi k_F^2E_F } \left[
{\omega^2_q} + {(2\pi T)^2 }\right]\ln{\bigl[\frac{\kappa
v_F}{\sqrt{\omega_q \Delta_q}}\bigr]},
\end{equation}
where $\Delta_q=\max(\omega_q,2\pi T)$. The expression
(\ref{omega}) weakly depends on the strength of the
electron-electron interaction. This is due to the fact that the
principal contribution to the conductivity comes, with logarithmic
accuracy, from the momentum range $Q<\kappa$, where interaction is
screened. Our calculation is, therefore, correct as long as the
Fermi momentum exceeds the inverse screening radius,
$k_F\gg\kappa$. In a realistic 2DEG, however, the two quantities
are of the same order of magnitude. In this general case it is no
longer a good approximation to disregard the scattering with large
momenta $Q\sim k_F$ and to neglect the exchange effects
represented by the amplitude ${\cal A}_{\rm ex}$ in
Eq.~(\ref{dissipnew}). One should expect the numerical coefficient
in Eqs.~(\ref{twopaircon})--(\ref{omega}) to change and become
interaction-dependent for $\kappa \gtrsim k_F$.

\section{Numerical estimates for a $\text{GaAs}$ heterostructure}

In this Section we compare the effectiveness of different
mechanisms contributing to the attenuation of plasmons. We
concentrate on the case of low temperature, and consider
attenuation $\gamma_q$ as a function of the plasmon energy
$\hbar\omega_q$.

In the important case of a GaAs heterostructure, the material
parameters are\cite{MPH,Z,B}: deformation potential coupling
constant $\Xi=2.2\times 10^{-18}$ J, piezoelectric constant
$h_{14}= 1.38\times 10^9$ V m$^{-1}$, longitudinal sound velocity
$s_l=5.2\times 10^3$ m~s$^{-1}$, transverse sound velocity
$s_t=3.0 \times 10^3$ m~s$^{-1}$, crystal density $\rho=5.3\times
10^3$ kg~m$^{-3}$, and the dielectric constant
$\varepsilon^*=12.8$. For estimates, we take a typical Fermi
momentum, $k_F \approx 1\times 10^8$ m$^{-1}$, which corresponds
to the Fermi energy $E_F \approx 5.6$ meV.

Using Eqs.~(\ref{coupling_d}) and (\ref{piezoconstant}) we find
the dimensionless electron-phonon coupling strengths: $\eta_d
=0.03$ for the deformation potential, and $\eta_p =0.09$ for the
piezoelectric coupling. The characteristic frequencies $\omega_s$
and  $\omega_\kappa$, introduced after Eqs.~(\ref{homowidth_d})
and (\ref{gammapiezo2}) respectively, are of the same order,
$\hbar\omega_\kappa\approx 0.7$ meV, and $\hbar\omega_s\approx
0.3$ meV.

Comparison of
Eqs.~(\ref{homowidth_d}),~(\ref{def2}),~(\ref{gammapiezo2}), and
(\ref{omega}) with each other shows that the plasmon-phonon
conversion mechanism yielding $\gamma_q\propto\omega_q^4$, see
Eq.~(\ref{homowidth_d}), dominates the attenuation only at very
low plasmon energies, $\hbar\omega_q\lesssim 0.02$ meV. At higher
energies, the phonon-assisted Landau damping via piezoelectric
coupling prevails, and $\gamma_q\propto\omega_q^5$ as long as
$\hbar\omega_q\lesssim 2\omega_s\approx 0.7$ meV. At higher
energies, the piezoelectric contribution saturates, see
Eq.~(\ref{gammapiezo1}), while the contribution coming from the
deformation potential interaction, Eq.~(\ref{def1}), monotonically
increases, $\gamma_q\propto\omega_q^2$. It becomes the dominant
one at $\hbar\omega_q\gtrsim 1$ meV, and exceeds the two-pair
mechanism, Eq.~(\ref{omega}), by a factor $\sim (E_F/\omega_q)^4$
in the entire range $\hbar\omega_q\lesssim E_F$. We are not
considering higher plasmon energies, where the conventional Landau
damping significantly contributes to the plasmon line width.

\section{Conclusions}

We presented a theory for the real part of the optical conductivity
$\sigma'(\omega,q)$ of a ballistic 2DEG at finite wave-vectors and
utilized it for calculation of a plasmon width.  Collisionless energy
dissipation (Landau damping) in a degenerate plasma is exponentially
small $\sim \exp{[-m\omega_q^2/2q^2T]}$ in the low-temperature and
long-wavelength limit. That prompted us to consider a number of
mechanisms which would be sub-leading, should conservation laws allow
for the Landau damping. These mechanisms are the plasmon-phonon
conversion, see Section II, phonon-assisted creation of an
electron-hole pair, see Section III, and two-pair absorption, see
Section IV. In the domain of low plasmon energies and low
temperatures, $\hbar\omega_q, T\lesssim E_F$, all the found
contributions have power-law (in $\omega_q$ and $T$) asymptotes, see
Eqs.~(\ref{homowidth_d}), (\ref{def2}), (\ref{gammapiezo2}),
(\ref{gammapiezo1}), (\ref{gammapiezo3}), and (\ref{omega}). The
smallest exponent belongs to the conversion mechanism, which therefore
dominates the plasmon line width at the lowest temperatures. (We do
not discuss here a competing contribution from the impurity
scattering.) The relative importance of the other considered
mechanisms at higher temperatures depends on the material parameters.
In the important case of a GaAs heterostructure, apparently the
two-pair mechanism does not become the leading one, see Section V.
This mechanism, however, results in a ``universal'' contribution, see
Eq.~(\ref{omega}), which is only weakly (logarithmically) dependent on
the parameters of the host material.

\section{Acknowledgements}

We acknowledge fruitful discussions with A.\ Andreev, E.\ Demler,
B.\ Halperin, A.\ Kamenev, D.\ Maslov, P.\ Platzman, M.\ Pustilnik
and V.\ Vinokur. E.M.\ is grateful to the University of Minnesota
for hospitality, and M.R.\ thanks Aspen Center for Physics. The
research at Harvard University was supported by NSF grant
DMR02-33773, and at the University of Minnesota by NSF grants
DMR02-37296 and EIA02-10736. E.M.\ also acknowledges the support
from the Russian Foundation for Basic Research, project \#
04-02-17087.

\appendix

\section{Plasmon-phonon conversion rate in the Golden rule
approximation}

The plasmon-phonon conversion rate, Eq.~(\ref{homowidth}), can
also be derived from the Golden rule formalism similar to the
phonon-assisted Landau damping and the two-pair attenuation rate.
The probability of electric field absorption accompanied by the
phonon emission, Fig.\ 4, is given by the expression,
\begin{eqnarray}
dw_{\lambda \bf k}=2\pi e^2 |\phi_0|^2 \chi^2(\omega,q)
\frac{{\cal M}^2_{\lambda}({\bf k})}{2\rho \omega_{\lambda k}}~
\delta(\omega -\omega_{\lambda k}) \nonumber\\ \times \delta({\bf
q}-{\bf k}_\parallel) \frac{d^3k}{2\pi}.
\end{eqnarray}
\begin{figure}[h]
\label{1} \resizebox{.20\textwidth}{!}{\includegraphics{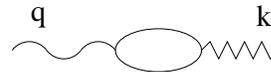}}
\caption{Transition amplitude for the direct conversion of the
electric field (wavy line) to a phonon (zigzag line). The loop
stands for the electron polarization operator $\chi(\omega,q)$.}
\end{figure}
After summation over phonon states, the total probability of the
field absorption becomes
\begin{equation}
\label{conv} I_a=\sum_\lambda \int (1+N_{\lambda k}) ~ dw_{\lambda
\bf k} .
\end{equation}
The total probability of the field emission (accompanied by the
phonon annihilation), $I_e$, is obtained from Eq.~(\ref{conv}) by
changing $1+N_{\lambda k} \to N_{\lambda k}$. As always, the two
probabilities are related by the detailed balance condition,
$I_e=I_a e^{-\omega/T}$. The corresponding contribution to the
optical conductivity is found from Eqs.~(\ref{dissipphon1}) and
(\ref{sigma}),
\begin{equation}
\sigma'(\omega,q)=\frac{e^2 \chi^2(\omega,q)}{2\rho
q^2}\sum_\lambda s_\lambda^{-1} {\cal
M}^2_\lambda(q,\omega/s_\lambda).
\end{equation}
Finally, the plasmon attenuation is determined by
Eq.~(\ref{gammaq}),
\begin{equation}
\gamma_q=\frac{\chi(\omega_q,q)}{4\rho}\sum_\lambda s_\lambda^{-1}
{\cal M}^2_\lambda(q,\omega_q/s_\lambda),
\end{equation}
which is equivalent to Eq.~(\ref{homowidth}).

\end{document}